\documentclass[twocolumn,pra,showpacs,amsmath,amssymb]{revtex4}

\usepackage{amsmath,amsfonts,amsthm,graphicx,color}

\usepackage{float}
\usepackage{hyperref}

\begin{document}

\title{More efficient Bell inequalities for Werner states}
\author{T. V\'ertesi}
\email{tvertesi@dtp.atomki.hu}
\affiliation{Institute of Nuclear Research of the Hungarian Academy of Sciences\\
H-4001 Debrecen, P.O. Box 51, Hungary}

\def\CC{\mathbb{C}}
\def\RR{\mathbb{R}}
\def\one{\leavevmode\hbox{\small1\normalsize\kern-.33em1}}
\newcommand*{\tr}{\mathsf{Tr}}
\newtheorem{theorem}{Theorem}[section]
\newtheorem{lemma}[theorem]{Lemma}

\date{\today}

\begin{abstract}
In this paper we study the nonlocal properties of two-qubit Werner states parameterized by the visibility parameter $0\le p\le 1$. New family of Bell inequalities are constructed which prove the two-qubit Werner states to be nonlocal for the parameter range $0.7056<p\le 1$. This is slightly wider than the range $0.7071<p\le 1$, corresponding to the violation of the Clauser-Horne-Shimony-Holt (CHSH) inequality. This answers a question posed by Gisin in the positive, i.e., there exist Bell inequalities which are more efficient than the CHSH inequality in the sense that they are violated by a wider range of two-qubit Werner states.
\end{abstract}

\maketitle

\section{Introduction}\label{intro}

Quantum Mechanics is inherently nonlocal, clearly demonstrated by the fact that measurements on quantum states may violate the so-called Bell inequalities \cite{Bell64,CHSH69}. This has been verified experimentally as well, up to some technical loopholes \cite{Aspect}.
On the other hand, when a quantum state cannot be prepared using only local operations and classical communication, it possesses quantum correlations and we say that the state is entangled. It was Werner who asked firstly what the relation is between quantum nonlocality and quantum correlations \cite{Werner}.
It is actually known that any pure entangled state of two or more subsystems may violate a generalized Bell inequality \cite{GP92,PR92}, thus here nonlocality and entanglement coincide. For mixed states, however, the relation between entanglement and nonlocality is much complicated. In 1989 Werner \cite{Werner} constructed a family of bipartite mixed states (became known as Werner states), which, while being entangled, yield outcomes that admit a local hidden variables (LHV) model. This conclusively proved that entanglement and nonlocality are different resources.

However, if we want to describe quantitatively the difference, the picture turns out to be quite subtle even in the case of two-qubit Werner states, which are mixtures of the singlet $|\psi^-\rangle=(|01\rangle - |10\rangle)/\sqrt 2$ with white noise of the form
\begin{equation}
\rho^W_p = p|\psi^-\rangle\langle \psi^-| + (1-p){\one}/4.
\label{werner}
\end{equation}
Werner showed \cite{Werner} that these states are separable if and only if $p\le 1/3$. With respect to the locality properties, on one hand, Werner states admit a LHV model for all measurements for $p\le 5/12$ \cite{Barrett} and admit a LHV model for projective measurements for $p\le0.6595$ \cite{AGT06}. On the other hand, Werner states violate the CHSH inequality for $p>1/\sqrt 2$, in which case LHV model clearly cannot be constructed. It is not known whether Werner states admit an LHV model in the region $0.6595<p\le 1/\sqrt 2$.
The actual value of $p$ where the state ceases to be nonlocal, designated $p_c^W$, is particularly relevant from the viewpoint of experiments since this value specifies the amount of noise the singlet tolerates before losing its nonlocal properties. This issue was addressed by Gisin some time ago \cite{open19} (see also \cite{Gisin07}), who posed the question to find Bell inequalities which are more efficient than the CHSH one for Werner states. In this paper we intend to give a definite answer to this question by providing Bell inequalities which can be violated slightly stronger than the CHSH one, resulting in the bound $p_c^W\le 0.7056$ for the nonlocality visibility threshold (instead of the bound $p_c^W\le 1/\sqrt 2\sim0.7071$ owing to the CHSH inequality). Note, how powerful the CHSH inequality is, which is the simplest Bell inequality, consisting of two settings on each side, while a stronger Bell inequality presented in this work has at least 465 settings on each side.

We also would like to point out that while in certain cases using sequence of measurements may extend the range of locality \cite{Popescu95}, the nonlocality threshold $p_c^W$ for Werner states could not be decreased even on this way \cite{Gisin96}.
There is also an interesting line of research, which explores the parameter region of Bell violation for Werner states by restricting the class of possible LHV models \cite{Pitowsky08,WD07}. Actually, Ref.~\cite{WD07} could achieve violation of certain Bell inequalities, assuming the above limitations for $p\ge 1/3$, i.e., for the entire range of the nonseparability region. An other way of generalization to obtain the range of locality is the extension of the Werner states to e.g., more parties \cite{TA06} or higher dimensions \cite{APBTA, WJD07}. However, let us mention, that a gap also remained in these cases between the best known local model \cite{APBTA, WJD07} and the proven nonlocality threshold \cite{CGLMP,KGZMZ}.

The outline of the present work is as follows. In Sec.~\ref{bellgroth} we briefly summarize the relation between Bell inequalities for two-qubit Werner states and Grothendieck constant of order 3, denoted by $K_G(3)$. In Sec.~\ref{cons} a family of Bell inequalities is constructed  and in Sec.~\ref{lower1} with the aid of these inequalities a lower bound, bigger than $\sqrt 2$, is given for $K_G(3)$, implying that Werner states (\ref{werner}) with $p<1/\sqrt 2$ can still violate these inequalities. In Sec.~\ref{lower2} we give a better lower bound for $K_G(3)$ and for higher orders ($K_G(d)$ with $d=4,5$), as well. In Sec.~\ref{other} a Bell inequality is provided with a number of settings 11 and 14, proving $K_G(4)>\sqrt 2$, and in Sec.~\ref{tight} the relevance property of the constructed family of Bell inequalities is demonstrated. Sec.~\ref{disc} summarizes and poses some open questions.

\section{Bell inequalities linked to Grothendieck constants}\label{bellgroth}

Define the expression
\begin{equation}
I=|\sum_{i,j=1}^m {M_{ij}\,a_i b_j}|,
\label{LHV}
\end{equation}
where $M$ is any $m \times m$ matrix with real entries and $a_1,\ldots,a_m,b_1,\ldots,b_m \in \{-1,+1\}$.
Now let us define
\begin{equation}
I^d=|\sum_{i,j=1}^m {M_{ij}\,\vec a_i\cdot \vec b_j}|,
\label{quantum}
\end{equation}
where the unit vectors $\vec a_1,\ldots,\vec a_m,\vec b_1,\ldots,\vec b_m$ are in $\RR^d$ and $\vec a\cdot\vec b$ is the dot product of $\vec a$ and $\vec b$.
Grothendieck constant plays a prominent role in the theorem of linear operators on Banach spaces \cite{Pisier}.
Grothendieck constant of order $d$, designated $K_G(d)$, for any integer $d \ge 2$, can be defined as \cite{Finch}
\begin{equation}\label{groth}
I^d\leq K_G(d) \max_{a_i,b_j=\pm 1} I
\end{equation}
for all unit vectors $\vec a_1,\ldots,\vec a_m,\vec b_1,\ldots,\vec b_m$ in $\RR^d$ and for all $m\times m$ matrix $M$.
The constant $K_G(d)$ is taken to be the smallest possible one.

Now let us discuss briefly the connection with Bell inequalities. In the Bell scenario we consider two parties, Alice and Bob, each chooses from $m$ $\pm 1$-valued observables, specified by $\{A_1,\ldots,A_m\}$ and $\{B_1,\ldots,B_m\}$. The joint correlation of Alice and Bob's measurement outcomes, designated $\alpha_i$ and $\beta_j$ respectively, is given by $\langle \alpha_i \beta_j\rangle=\tr(A_i\otimes B_j\rho)$, where $\rho$ denotes the density matrix of the bipartite state. A correlation Bell inequality can be written as
\begin{equation}
\sum_{i,j=1}^m{M_{ij}\langle\alpha_i \beta_j\rangle}\le L,
\label{bell}
\end{equation}
where $L$ signifies the bound which can be achieved by local models and $M$ is a $m\times m$ matrix with real coefficients defining a Bell inequality. The local bound can always be achieved by a deterministic local model, i.e., for all real numbers $a_i,b_j=\pm 1$ we have
\begin{equation}
\max_{a_i,b_j}{\sum_{i,j=1}^m{M_{ij}a_i b_j}}= L.
\end{equation}
In this way the expression $I$ defined by (\ref{LHV}) is linked to a correlation Bell inequality with matrix $M$ and local bound $\max_{a_i,b_j=\pm 1}{I}=L$.

On the other hand, for the singlet state $\rho=|\psi^-\rangle\langle\psi^-|$ we have $\langle \alpha_i \beta_j \rangle_{\psi^-} = \langle\psi^-|A_i\otimes B_j|\psi^-\rangle= -\vec a_i\cdot\vec b_j$, where the observables $A=\vec a \vec\sigma$ and $B=\vec b \vec\sigma$ corresponding to Alice and Bob's projective measurements are specified by the unit vectors $\vec a$ and $\vec b$ in $\RR^3$. Then substituting into (\ref{bell}) one obtains the expression $I^3$ in (\ref{quantum}). Furthermore, Tsirelson \cite{Tsirelson87} proved that correlations which are dot products of unit vectors $\vec a, \vec b\in\RR^d$ can always be realized by performing projective measurements on maximally entangled states in some higher dimensional Hilbert spaces. Thus the value $\max I^d$ can always be achieved by means of quantum mechanics.

Since joint correlations vanish for the maximally mixed state, it follows that the critical point at which Werner states in (\ref{werner}) cease to violate any Bell inequality is $p_c^W = 1/K_G(3)$. This key correspondence has been established in Ref.~\cite{AGT06}. Though, the exact value of $K_G(3)$ is not known, but known bounds establish that $0.6595 \le p_c^W \le 0.7071$. In this paper we show that $K_G(3)\ge 1.4172$ implying the slightly smaller gap $0.6595 \le p_c^W \le 0.7056$.
Let us mention, that the Fishburn-Reeds Bell inequality \cite{FR94} provides an explicit example
with 20 settings on each side showing that $K_G(5)\ge 10/7=1.42857>\sqrt 2$. Also Toner has shown that $K_G(4)>\sqrt 2$ \cite{AGT06}. But, as far as we know, the question has remained open whether $K_G(3)$ is bigger than $\sqrt 2$ implying $p_c^W<1/\sqrt 2$.

\section{Constructing family of Bell inequalities}\label{cons}

For Bell diagonal states, such as for Werner states, under projective measurements Alice and Bob's local marginals (defined by $\langle \alpha_i\rangle =\tr(A_i\otimes\one\rho)$ for Alice and likewise for Bob) are zero, thus it is sufficient to deal with generic correlation Bell inequalities defined by (\ref{bell}) to obtain maximal Bell violation for Werner states. Moreover, in this respect, the tight correlation Bell inequalities, which can be considered as facets of the correlation polytope \cite{Pitowsky89}, specified by the number of two-outcome measurements $m$ on each side, are the most efficient ones. For $m=2$ one obtains as the only nontrivial correlation inequality the CHSH one \cite{Tsirelson80}. For $m>2$ one needs to resort to numerical programs for computing the inequalities corresponding to the inequivalent facets of the correlation polytope. Up to $m=4$ all the correlation inequalities have been computed \cite{AII06}, and the two inequivalent inequalities obtained are in fact less efficient than the CHSH one for Werner states. However, the complexity of the computation exponentially grows with $m$ (in fact, this is an NP-complete problem \cite{Pitowsky89}), therefore there is no hope to completely characterize all the facets of the correlation polytope for any given $m$. Thus in general one needs to look for alternative methods. For instance Gisin explored special form of families of tight correlation inequalities, the so-called $D$-inequalities in Ref.~\cite{Gisin07}. Avis et al.~ \cite{AII06} applied triangular elimination to the list of known facet inequalities of the cut polytope to construct many new tight correlation inequalities. Alternatively, one can construct (possibly not tight) correlation inequalities which however can be easily generalized to arbitrary number of settings, such as in the cases \cite{BC90,Gisin99,BG03,VP08}. In the present work we have chosen this latter direction by modifying the correlation inequalities $Z_n$ introduced in \cite{VP08}.

Let us specify the form of $M$ in (\ref{LHV}) through the following formula,
\begin{align}
I_{n_A,n_B} =& \sum_{i=1}^{n_A}\sum_{j=1}^{n_B}{a_i b_j} \nonumber\\&+ \sum_{1\le i<j\le n_B}{a_{ij}(b_i-b_j)} + \sum_{1\le i<j\le n_A}{b_{ij}(a_i-a_j)},
\label{Inanb}
\end{align}
entailing a Bell inequality with $m_A=n_A+n_B(n_B-1)/2$ and $m_B=n_B+n_A(n_A-1)/2$ measurement settings on Alice and Bob's respective side. First we calculate the maximum achievable value (local bound) for it. For this sake, we can write for the maximum
\begin{align}
\max I_{n_A,n_B} =& \max_{a_i,b_j=\pm 1} \{\sum_{i=1}^{n_A}{a_i}\sum_{j=1}^{n_B}{b_j} \nonumber\\&+ \sum_{1\le i<j\le n_B}{|b_i-b_j|} + \sum_{1\le i<j\le n_A}|a_i-a_j|\}.
\label{maxInn}
\end{align}
Generally, the local bound can be obtained by finding the maximum over all possible values $a_i,b_j=\pm 1$. However, in this particular case one can exploit the symmetry with respect to change of indices within the sets $\{a_i\}_{i=1}^{n_A}$ and $\{b_i\}_{i=1}^{n_B}$. Thus one needs to check altogether $n_A n_B$ cases where $+1$ occurs $1\le k\le n_A$ times in the set $\{a_i\}_{i=1}^{n_A}$ and $+1$ occurs $1\le l\le n_B$ times in the set $\{b\}_{i=1}^{n_B}$ (the rest being $-1$).
For any $k,l$ pair we have $\max{I_{n_A,n_B}}=\max\{(n_A-2k)(n_B-2l)+2(n_A-k)k+2(n_B-l)l\}$. This expression is maximal by $k-l=\lfloor (n_A-n_B)/2 \rfloor$ resulting in the local bound
\begin{align}
\max{I_{n_A,n_B}}=&(n_A^2+n_B^2-1)/2,\;\; \mathrm{for}\;|n_A-n_B|\; \mathrm{odd} \nonumber\\
\max{I_{n_A,n_B}}=&(n_A^2+n_B^2)/2,\;\;   \mathrm{for}\;|n_A-n_B|\; \mathrm{even}.
\label{localmax}
\end{align}

In this paper we focus on two particular cases $n_A=n_B+1$ and $n_A=n_B$, but first let us restrict our attention to the latter, symmetric case $n_A=n_B=n$. The LHV bound gives $\max I_{n,n}=n^2$ by inserting $n_A=n_B=n$ in (\ref{localmax}).
On the other hand, the expression $I^d_{n,n}$, symmetric in the two parties, reads
\begin{align}
I^d_{n,n}=&\sum_i^{n}\sum_j^{n}{\vec a_i \cdot \vec b_j} \nonumber\\ &+ \sum_{1\le i<j\le n}{\vec a_{ij}\cdot(\vec b_i-\vec b_j)} + \sum_{1\le i<j\le n}{\vec b_{ij}\cdot(\vec a_i-\vec a_j)}.
\label{Idnn}
\end{align}
For the maximum, similarly to the LHV case, the two-indices terms can be omitted:
\begin{align}
&\max I^d_{n,n}= \max\{\sum_{i=1}^n{\vec a_i}\sum_{j=1}^n{\vec b_j} + \sum_{1\le i<j\le n}{|\vec b_i- \vec b_j|} \nonumber\\&+ \sum_{1\le i<j\le n}|\vec a_i-\vec a_j|\} \le \max\{\frac{1}{2}|\sum_{i=1}^n{\vec a_i}|^2 + \frac{1}{2}|\sum_{i=1}^n{\vec b_i}|^2 \nonumber\\&+ \sum_{1\le i<j\le n}{|\vec a_i-\vec a_j|} + \sum_{1\le i<j\le n}{|\vec b_i-\vec b_j|} \},
\label{maxIdnn}
\end{align}
where the maximization is over all $\vec a_i,\vec b_j\in S^{d-1}$ and the last inequality comes from the relation between the geometric and quadratic mean. Furthermore, since $\{\vec a_i\}_{i=1}^{n}$ and $\{\vec b_i\}_{i=1}^{n}$ do not depend on each other, one can maximize the two sets independently resulting in
\begin{equation}
\max I^d_{n,n}=\max\{|\sum_{i=1}^n{\vec a_i}|^2 + 2\sum_{1\le i<j\le n}|\vec a_i-\vec a_j|\},
\label{maxIdnn2}
\end{equation}
with the constraints $\vec a_i\in S^{d-1}$, where the equality sign is due to the fact that at the maximum one can take $\vec a_i=\vec b_i$ for all $i$, which saturates the inequality in (\ref{maxIdnn}). This expression shows some similarity with the one appearing in \cite{VP08}, in which case one had to maximize only the last term, i.e., the sum of distances of $n$ unit vectors. This is a problem occurring in discrete mathematics, and there exist optimal solutions for various instances of the values $n,d$ \cite{CK07}. In contrast, in the present case, the quadratic term makes things complicated to get the true optimum value. However, due to formula~(\ref{groth}) one can give the lower bound $K_G(d)\ge I_{n_A,n_B}^d/ \max{I_{n_A,n_B}}$ for the Grothendieck constant of order d, without knowing the true maximum value $\max{I_{n_A,n_B}^d}$.

\section{Lower bound for Grothendieck constant of order 3}\label{lower1}

In fact, in the particular case $I_{n,n}^n$ one can obtain the exact maximum, $\max{I_{n,n}^n}=3/2-1/(2n)$. This result (noticing that for large $n$ the violation tends to $1.5$) would indicate that there may be some hope to get a lower bound $K_G(3)> \sqrt 2$. Below we show that the maximum above can indeed be attained.

First let us observe that in (\ref{maxIdnn2}) only $n$ vectors occur, thus we have $\max I^m_{n,n}=\max I^n_{n,n}$ with $m=n(n+1)/2$, where $m$ is the number of measurement settings on each side. Now we take the unit vectors $\vec a_i$ in such a way that $\vec a_i \cdot \vec a_j = 1/2$ for all $i\ne j$. This can be achieved, by noting that the $n\times n$ Gram matrix $G$, defined by elements $G_{ij}=\vec a_i\cdot\vec a_j$ is positive definite, and every positive definite matrix is a Gram matrix for some set of vectors. Thus it is enough to show that the Gram matrix $G$, defined as above ($G_{ij}= 1,\, \forall i=j$ and $G_{ij}=1/2,\, \forall i\ne j$), is positive definite. However, using the Sylvester criterion \cite{HJ} one can establish that $G$ defined above is positive definite iff $\det G>0$ for any $n$. One may obtain by induction the closed formula $\det A = (a-b)^{n-1}(a+(n-1)b)$ for the determinant of any $n\times n$ matrix $A$ having in the diagonals the value $a$ and in all off-diagonals the value $b$. By choosing particularly $a=1$ and $b=1/2$ one gets $\det G>0$ for any dimension $n$, which proves our assertion.

Beside, all the elements in (\ref{maxIdnn2}) can be obtained as the only function of $\vec a_i\cdot\vec a_j$,
since $|\sum_{i=1}^n \vec a_i|^2=\sum_{i,j=1}^n \vec a_i \cdot \vec a_j$ and $|\vec a_i - \vec a_j|=\sqrt{2-2\vec a_i\cdot\vec a_j}$. Thus by substitution we obtain $I^n_{n,n}=n(3n-1)/2$. Then it follows using (\ref{localmax}), that $I^n_{n,n}/\max{I_{n,n}} = 3/2 - 1/(2n)$. It is also possible to verify that this is in fact the maximum value.
The verification, which is not detailed here, goes the same line as discussed by Wehner in \cite{Wehner06} for the chained Bell inequality \cite{BC90} through the dual solution of a semi-definite optimization problem \cite{BV04}. Note that this optimization problem is just the first step in the hierarchy introduced by Navascues et al. \cite{NPA07,NPA08}.

Now we wish to obtain a lower bound $K_G(3)\ge I_{n,n}^3/\max{I_{n,n}}$ bigger than $\sqrt 2$ for the Grothendieck constant of order 3, $K_G(3)$. Since owing to (\ref{localmax}), $\max I_{n,n}=n^2$, we are left with the calculation of $I_{n,n}^3$ which though might be not maximal, but large enough to supply a good lower bound for $K_G(3)$.
This is achieved by substituting in (\ref{maxIdnn2}) in the place of $\vec a_i$ explicit values on the following way. Since $\vec a_i$ are unit vectors in $\RR^3$, the first two components $P_i=(x_i,y_i)$ of $\vec a_i$, which can be considered as points in the $XY$-plane, completely specify the vector itself. Let $n=30$ and distribute these points on three co-centric circles centered at the origin $(0,0)$ with radii $\rho_I=22/100$, $\rho_{II}=52/100$ and $\rho_{III}=77/100$. Then let $P_1=(\rho_I\cos \pi/4, \rho_I\sin \pi/4)$, $P_5=(0, \rho_{II})$ and $P_{15}=(0, \rho_{III})$. The other $P_i$ vectors are constructed from the above vectors by rotating them with angles $\pi/2$, $\pi/5$, and $\pi/8$, respectively, such as to form regular polygons with vertices 4, 10 and 16 (as it is shown in the figure).
By inserting the explicit values of the corresponding set $\{\vec a_i\}_{i=1}^{30}$ into the expression to be maximized in (\ref{maxIdnn2}) one obtains $I_{n,n}^3/I_{n,n}=1.415~199$ with $n=30$. This implies that $K_G(3)$ is indeed bigger than $\sqrt 2=1.414~2136\ldots$.
The specific values of $\vec a_i$ have been found by performing optimization with respect to the radii of the three circles by choosing regular polygons with various number of vertices.

\begin{figure}
\includegraphics[width=8cm]{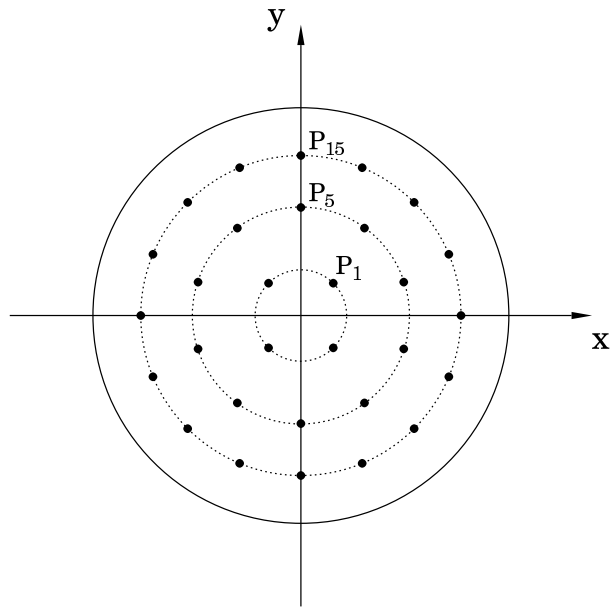}
\caption{
The 30 points which are projection of the vectors $\vec a_i$ on the $XY$-plane.
They are equally distributed on three co-centric circles with radii $22/100, 52/100, 77/100$ centered at the origin.
The outer circle represents the grand circle projected on the $XY$-plane, thus having radius 1.
}
\label{fig-circles}\end{figure}

\section{Better lower bounds for Grothendieck constant of orders 3,4,5}\label{lower2}

In this section a general method is discussed to obtain local maximum on $I_{n,n}^d$ for any $n,d$, which for many instances are presumably the global or close to the global maximum. Then, recalling $K_G(d)\ge I_{n,n}^d/ \max{I_{n,n}}$ and $\max{I_{n,n}}=n^2$, this method yields lower bounds for $K_G(d)$. In particular we present results for $d=3,4,5$, calculated by the value $n=100$.

Let us consider the following iteration scheme, which is a simplified version of the see-saw iteration method, already used in the literature to solve optimization problems in similar context entering many optimization parameters \cite{WW01,IID06}.
Note, that the matrix $M$ of $I_{n,n}$ defined through (\ref{Inanb}) is symmetric, thus we may write
\begin{equation}
I_{n,n}^d=\sum_{i,j=1}^m{M_{ij}\vec a_i\cdot \vec b_j}=\sum_{i=1}^m{\vec b_i\cdot\sum_{j=1}^m{M_{ij}\vec a_j}}=\sum_{i=1}^m{\vec a_i\cdot\sum_{j=1}^m{M_{ij}\vec b_j}},
\label{seesaw}
\end{equation}
with $m=n(n+1)/2$ and $\vec a_i,\vec b_j$ are unit vectors in $\RR^d$. In this notation we contracted the double indices $ij$ appearing in (\ref{Idnn}), so that $\{\vec a_i\}_{i=1}^m$ stands for the set $(\{\vec a_i\}_{i=1}^n,\{\vec a_{ij}\}_{1\le i<j\le n})$ and similarly for the vectors $\vec b$.

Considering (\ref{seesaw}) one can maximize the expression $I_{n,n}^d$ for given $\{\vec a_i\}_{i=1}^m$ by setting
$\vec b_i$ parallel to $\sum M_{ij}\vec a_j$ for all $i$. Then one can continue with setting $\vec a_i$ parallel to $\sum M_{ij}\vec b_j$ for all $i$. However, due to the fact that $M$ is symmetric, one can get rid of the vectors $\vec b$ and  obtain the iteration rule $\vec a_i \rightarrow \sum_j{M_{ij}\vec a_j}/|\sum_j{M_{ij}\vec a_j}|$ for all $i$, provided $|\sum_j{M_{ij}\vec a_j}|\ne 0$. Here the notation $|\vec v|$ refers to the Euclidean norm of a vector $\vec v\in\RR^d$. Thus our task is to give initial values for the unit vectors $\vec a_i\in\RR^d$, which we choose in the following way.
The surface of the unit sphere in $\RR^d$ can be parameterized by $d-1$ angles, $\vec a_i=(\cos(\phi^1_i),\sin(\phi^1_i)\cos(\phi^2_i),\ldots, \sin(\phi^1_i)\cdots\sin(\phi^{d-2}_i)\cos(\phi^{d-1}_i),$ $\sin(\phi^1_i)\cdots\sin(\phi^{d-2}_i)\sin(\phi^{d-1}_i))$.
We define the starting vectors $\{\vec a_i\}_{i=1}^m$ with angles $\phi^{k}_i=k\,[\mathrm{rad}],\; 1\le k\le d-1$.
Then we perform the above iteration scheme for a given time, practically till the vectors $\vec a_i$ in two successive iteration steps differ less than an infinitesimal threshold value. In particular, for $n=100$ we found that $1000$ iteration steps were sufficient for our purposes. Also, we checked for each case $d=3,4,5$, that the value $|\sum_j{M_{ij}\vec a_j}|$ in the denominator of the iterated expression was nonzero (actually, it was no less than $10^{-4}$ for all $i$ in each case of $d$). On the other hand, the iteration was performed with machine precision $\sim10^{-16}$ in the Mathematica package, and we checked that $|\vec a_i\cdot\vec a_i -1|<10^{-15}$ for all $1\le i\le m$ after the $1000$ iteration steps completed.

For $n=100$, we obtained the following numbers, $I_{n,n}^3/\max I_{n,n}=1.417~241$, $I_{n,n}^4/\max I_{n,n}=1.445~207$, and $I_{n,n}^5/\max I_{n,n}=1.460~065$. These numbers are lower bounds for the Grothendieck constants $K_G(3)$, $K_G(4)$ and $K_G(5)$, respectively. We mention that for $n=100$ the dimension of the respective matrix $M$ is $n(n+1)/2=5050$. Note that the best lower bound for $K_G(5)$ presented so far in the literature $K_G(5)\ge 10/7= 1.428~571\ldots$ comes from the Fishburn-Reeds inequality \cite{FR94}. Our result for $K_G(3)$ provides us with the better lower bound $p_c^W\le 0.705~596$ for the critical value $p_c^W$ owing to the formula $p_c^W=1/K_G(3)$.

\section{Minimal number of measurements}\label{other}

One may also ask, what is the smallest number of settings on Alice and Bob's side, where $K_G(d)$ can exceed $\sqrt 2$ for some $d>2$. To the best of our knowledge, so far it has been provided by the Fishburn-Reeds inequalities \cite{FR94}. Their construction, giving $K_G(5)\ge 10/7= 1.428~571\ldots$ can be obtained by 20 measurement settings on each side.
Now we choose $n_A=n_B+1=5$ in expression $I_{n_A,n_B}$ of (\ref{Inanb}), giving the number of settings 11 and 14 on Alice and Bob's side, respectively. Thus the matrix $M$ in this particular instance has dimensions $11\times 14$. We show that this expression $I_{5,4}$ provides us with an example where $K_G(4)>\sqrt 2$.

Substituting values $n_A=5$ and $n_B=4$ into the formula (\ref{localmax}) for odd $|n_A-n_B|$ one obtains the value 20 for the local bound. On the other hand, the maximum value corresponding to the vectorial case $\max{I_{5,4}^4}$, can be obtained by the mean of semidefinite techniques \cite{Wehner06} as a first step of the hierarchy in \cite{NPA07}, where we used the SeDuMi package \cite{sedumi} for Matlab by the explicit numerical computation. This algorithm solves both the primal and the dual optimization problem at the same time and thus yields bounds on the accuracy of the obtained solution as well. Actually, we obtained the same optimal value $28.390~139$ for both cases. This yields the ratio $1.419~507$ for the violation of the Bell inequality $I_{5,4}\le 20$ clearly beating the $\sqrt 2$ limit with 11 and 14 settings on Alice and Bob's side, respectively.

\section{Tightness and relevance of Bell inequalities}\label{tight}

It would be interesting to know whether the family of correlation inequalities defined by (\ref{Inanb}) is tight, i.e., whether it is a facet or not of the local Bell polytope \cite{Froissart} consisting of local marginals as well. This can be done by computing the dimension of the subspace spanned by all deterministic strategies saturating the inequality. If this subspace is found to be a hyperplane with dimension $d=m_A m_B+m_A+m_B$, then the inequality is tight.
Numerically, we treated the $n_A=n_B+1$ and $n_A=n_B=n$ cases in the expression $I_{n_A,n_B}$. Computationally we found that in the former case the inequality is tight up to $n_A=4$. On the other hand, the latter symmetric inequality proved to be not tight, but by the addition of some local terms $a_i,b_j=\pm 1$ as follows,
\begin{align}
I'_{n,n} =& \sum_{i,j=1}^n{a_i b_j} + \sum_{1\le i<j\le n}{a_{ij}(b_i-b_j)}\nonumber\\& + \sum_{1\le i<j\le n}{b_{ij}(a_i-a_j)}+\sum_{i=1}^n{a_i}-\sum_{j=1}^n{b_j},
\label{symmod}
\end{align}
we checked computationally its tightness up to $n=4$. Note, that here the terms $a_i,b_j$ refer to Alice and Bob's local marginals in the corresponding Bell inequalities.

Recalling from Sec.~\ref{cons}, that for $n_A=n_B$ we have $\max{I_{n,n}}=\max\{n^2-2(k-l)^2\}$, by adding the marginals $\max{I'_{n,n}}=\max\{n^2-2(k-l)^2+(2k-n)-(2l-n)\}=\max\{n^2-2(k-l)(k-l+1)\}=n^2$, thus the local bound does not change.
Let us notice, that $I'_{2,2}$ specified by $n=2$ in (\ref{symmod}) is just the $I_{3322}$ Bell inequality \cite{Froissart, CG04}, which is known to be tight \cite{AI05}. In both cases, $I_{n,n-1}$ and $I'_{n,n}$, we suspect that these Bell inequalities are tight for any higher values of $n>4$, as well.

Let us discuss the concept of relevant Bell inequalities, whose definition we quote from \cite{Gisin07}, Sec.~A.1: ``An inequality is relevant with respect to a given set of inequalities if there is a quantum state violating it, but not violating any of the inequalities in the set.'' Collins and Gisin \cite{CG04} showed that the $I_{3322}$ inequality is relevant to the famous CHSH inequality \cite{CHSH69}. Interestingly, they also found that given $I_{3322}$ the CHSH inequality is no longer relevant. Furthermore, Ito, Imai and Avis \cite{AII06} have recently conjectured supported by numerical optimization, that there exist Bell inequalities relevant for the $I_{3322}$ inequality for 3-level isotropic states. However, limiting the Hilbert space dimension to a qubit pair, they did not find a Bell inequality which would be relevant with respect to $I_{3322}$. Our new inequalities, $I_{n,n}$ and $I'_{n,n}$ with $n=100$, however, are examples to this latter case, demonstrating that in the parameter range $0.705~596 < p\le 0.7071~107$ two-qubit Werner states do not violate the $I_{3322}$ inequalities but violate $I_{n,n}$ or $I'_{n,n}$ for $n= 100$ (note that for the Werner states the local marginals become identically zero, thus in this respect $I_{n,n}$ and $I'_{n,n}$ are equivalent).

Moreover, one can demonstrate, that there is an inclusion relation, a notion introduced in \cite{AIIS05}, between $I'_{n,n}$ and $I'_{n-1,n-1}$, meaning that one can obtain the inequality $I'_{n-1,n-1}$ by measuring the identity for some settings in the inequality $I'_{n,n}$ (i.e., performing degenerate measurements). This implies that $I'_{n,n}$ for any $n>2$ is relevant with respect to $I'_{2,2}\equiv I_{3322}$. The proof is simple, actually by setting $a_n=+1, b_n=+1$ and $a_{in}=-1, b_{in}=-1$ for $1\le i<n$ in $I'_{n,n}$ one obtains $I'_{n-1,n-1}$, and then by induction one arrives at $I'_{2,2}$.

Altogether, one can say that if one limits the Hilbert space dimension to two qubits (\cite{Gisin07}, Sec.~A.2) the $I'_{n,n}$ inequality for $n\rightarrow\infty$ is the only relevant one with respect to all presently known Bell inequalities.

\section{Summary}\label{disc}

We provided a new family of Bell inequalities which proves that Werner states in (\ref{werner}) are nonlocal for the parameter range $p>0.7056$, the best earlier result $p>0.7071$ is given by the CHSH inequalities. Some of these Bell inequalities are shown to be relevant with respect to any other known Bell inequality. Our results have been obtained by proving that the Grothendieck constant of order 3, $K_G(3)$, is bigger than $\sqrt 2$, in particular, $K_G(3)\ge 1.4172$.
Though our result for the wider visibility range of nonlocal Werner states has been obtained for a number of settings (at least 465 for each party) which are not particularly suited for experiments, we believe that they are interesting from a conceptual point of view. Entangled states in many quantum information protocols (for instance in quantum communication complexity \cite{BZZ} and device-independent quantum key distribution problems \cite{quantumkey}) give advantage over there classical counterparts only if they exhibit nonlocal correlations. Thus, in this paper we have shown that this nonlocal correlation can in principle be exploited in a wider range of Werner states.

We leave it as an open question how to construct even better inequalities which would allow to beat the $\sqrt 2$ limit of $K_G(3)$ stronger. The possibility for such inequalities is suggested by the fact that an upper bound for $K_G\equiv \lim_{n\rightarrow \infty}K_G(n)$ is $1.7822$ which is suspected to be tight \cite{Krivine79,Finch}. But the inequality $I_{n,n}$ for $n\rightarrow\infty$ gives the lower bound $1.5$ for $K_G$, which is even smaller than the lower bound 1.6770 for $K_G$ presented in Ref.~\cite{DavieReeds}. Thus it is not impossible that there exist inequalities providing bigger values for $K_G(3)$ entailing even better Bell inequalities than the present ones for Werner states.


\begin{thebibliography}{9}

\bibitem{Bell64}
J.S. Bell, Physics {\bf 1}, 195 (1964).

\bibitem{CHSH69}
J. Clauser, M. Horne, A. Shimony, and R. Holt, Phys. Rev. Lett. {\bf 23}, 880 (1969).

\bibitem{Aspect}
A. Aspect, Nature {\bf 398}, 189 (1999).

\bibitem{Werner}
R.F. Werner, Phys. Rev. A {\bf 40}, 4277 (1989).

\bibitem{GP92}
N. Gisin and A. Peres, Phys. Lett. A {\bf 162}, 15 (1992).

\bibitem{PR92}
S. Popescu and D. Rohrlich, Phys. Lett. A {\bf 166}, 293 (1992).

\bibitem{Barrett}
J. Barrett, Phys. Rev. A {\bf 65}, 042302 (2002).

\bibitem{AGT06}
A. Acin, N. Gisin, and B. Toner, Phys. Rev. A, {\bf 73}, 062105 (2006).

\bibitem{open19}
N. Gisin, problem 19 (2003), presented in the web page \verb+http://www.imaph.tu-bs.de/qi/problems+.

\bibitem{Gisin07}
N. Gisin, arXiv:quant-ph/0702021v2 (2007).

\bibitem{Popescu95}
S. Popescu, Phys. Rev. Lett. {\bf 74}, 2619 (1995).

\bibitem{Gisin96}
N.Gisin, Phys. Lett. A {\bf 210}, 151 (1996).

\bibitem{Pitowsky08}
I. Pitowsky, J. Math. Phys. {\bf 49}, 012101 (2008).

\bibitem{WD07}
C.F. Wildfeuer and J.P. Dowling, arXiv:0708.1973v3 (2007).

\bibitem{TA06}
G. Toth and A. Acin, Phys. Rev. A {\bf 74}, 030306(R) (2006).

\bibitem{APBTA}
M.L. Almeida, S. Pironio, J. Barrett, G. Toth, and A. Acin, Phys. Rev. Lett. {\bf 99}, 040403 (2007).

\bibitem{WJD07}
H.M. Wiseman, S.J. Jones, and A.C. Doherty, Phys. Rev. Lett. {\bf 98}, 140402 (2007).

\bibitem{CGLMP}
D. Collins, N. Gisin, N. Linden, S. Massar, and S. Popescu, Phys. Rev. Lett. {\bf 88}, 040404 (2002).

\bibitem{KGZMZ}
D. Kaszlikowski, P. Gnacinski, M. Zukowski, W. Miklaszewski, and A. Zeilinger, Phys. Rev. Lett. {\bf 85}, 4418 (2000).

\bibitem{Pisier}
G. Pisier, {\it Factorization of Linear Operators and Geometry of Banach Spaces}, (American Mathematical Society, Providence, RI, 1986).

\bibitem{Finch}
S.R. Finch, {\it Mathematical Constants}, (Cambridge, Cambridge University Press, 2003), pp. 235-237.

\bibitem{Tsirelson87}
B.S. Tsirelson, J. Soviet. Math., {\bf 36}, 557, (1987).

\bibitem{FR94}
P.C. Fishburn and J.A. Reeds, SIAM J. Discrete Math. {\bf 7}, 48 (1994).

\bibitem{Pitowsky89}
I. Pitowsky, {\it Quantum Probability, Quantum Logic}, (Springer-Verlag, 1989).

\bibitem{Tsirelson80}
B.S. Tsirelson, Lett. Math. Phys., {\bf 4}, 93 (1980).

\bibitem{AII06}
D. Avis, H. Imai, T. Ito, J. Phys. A {\bf 39}, 11283 (2006).

\bibitem{BC90}
S. Braunstein and C. Caves, Ann. Phys. (NY), {\bf 202}, 22 (1990).

\bibitem{Gisin99}
N. Gisin, Phys. Lett. A, {\bf 260}, 1 (1999).

\bibitem{BG03}
H. Bechmann-Pasquinucci and N. Gisin, Phys. Rev. A {\bf 67}, 062310 (2003).

\bibitem{VP08}
T. Vertesi and K.F. Pal, Phys. Rev. A {\bf 77}, 042106 (2008).

\bibitem{CK07}
H. Cohn and A. Kumar, J. Am. Math. Soc., {\bf 20}, 99 (2007).

\bibitem{HJ}
R.A. Horn and C.R. Johnson, {\it Topics in Matrix Analysis}
(Cambridge University Press, Cambridge UK, 1991)

\bibitem{Wehner06}
S. Wehner, Phys. Rev. A, {\bf 73}, 022110 (2006).

\bibitem{BV04}
S. Boyd and L. Vandenberghe, {\it Convex Optimization}, (Cambridge University Press, 2004).

\bibitem{NPA07}
M. Navascues, S. Pironio, and A. Acin, Phys. Rev. Lett. {\bf 98}, 010401 (2007).

\bibitem{NPA08}
M. Navascues, S. Pironio, and A. Acin, arXiv:0803.4290 (2008).

\bibitem{WW01}
R. F. Werner and M. M. Wolf, Quantum Information \&
Computation {\bf 1}, 1 (2001), arXiv:quant-ph/0107093 (2001).

\bibitem{IID06}
T. Ito, H. Imai, D. Avis, Phys. Rev. A {\bf 73}, 042109 (2006).

\bibitem{sedumi}
J. Sturm, SeDuMi, a MATLAB toolbox for optimization over symmetric cones, URL
\verb+http://sedumi.mcmaster.ca+.

\bibitem{Froissart}
M. Froissart, Nuovo Cimento B {\bf 64}, 241 (1981).

\bibitem{CG04}
D. Collins and N. Gisin, J. Phys. A: Math. Gen. {\bf 37}, 1775 (2004).

\bibitem{AI05}
D. Avis and T. Ito, Discrete Appl. Math., {\bf 155}, 1689 (2007), arXiv:quant-ph/0505143 (2005).

\bibitem{AIIS05}
D. Avis, H. Imai, T. Ito, Y. Sasaki, J. Phys. A. {\bf 38} 10971 (2005), arXiv:quant-ph/0505060 (2005).

\bibitem{Krivine79}
J. Krivine, Adv. Math. {\bf 31}, 16 (1979).

\bibitem{DavieReeds}
A.M. Davie (unpublished note, 1984) and J.A. Reeds (unpublished note, 1991).

\bibitem{BZZ}
C. Brukner, M. Zukowski, and A. Zeilinger, Phys. Rev. Lett. {\bf 89}, 197901 (2002).

\bibitem{quantumkey}
J. Barrett, L. Hardy, and A. Kent, Phys. Rev. Lett. {\bf 95}, 010503 (2005);
A. Acin, N. Gisin, and L. Masanes, Phys. Rev. Lett. {\bf 97}, 120405 (2006);
A. Acin, N. Brunner, N. Gisin, S. Massar, S. Pironio, and V. Scarani,
Phys. Rev. Lett. {\bf 98}, 230501 (2007).

\end{thebibliography}
\end{document}